\documentstyle[aps,amsfonts,amssymb]{revtex}

\begin{document} 
\title{Anomalous oscillations of average transient lifetimes near  
crises }  
\author{Krzysztof Kacperski\cite{emkk}   
and Janusz A. Ho\l yst\cite{emjh} } 
\address{ Institute   of   Physics,   Warsaw  
University of Technology, Koszykowa 75,  
PL-00--662 Warsaw, Poland \\  and  \\
Max Planck Institute for Physics of Complex Systems,
N\"othnitzer Str. 38,
01187 Dersden, Germany
} 
\date{\today} 
\maketitle

\begin{abstract} 
 
It is common that the average length of chaotic transients appearing  
as a consequence of crises in dynamical systems obeys a power low of 
scaling with the distance from the crisis point. It is, however,  
only a rough trend; in some cases considerable oscillations can be 
superimposed on it. In this letter we report anomalous oscillations 
due to the intertwined structure of basins of attraction.  
We also present a simple geometrical model that gives an 
estimate of the period and amplitude of these oscillations.  
The results obtained  
within the model coincide with those yielded by computer  
simulations of a kicked spin model and the H\'enon map.

\end{abstract} 
 

PACS: 05.45.+b

Keywords: Crises; crisis-induced intermittency; chaotic transients;

\section{Introduction}
 
Among a large diversity of phenomena investigated in the scope of  
deterministic chaos in nonlinear dynamical systems one finds  
{\it crises} \cite{b4}. 
These are sudden changes in the structure of a chaotic attractor  
due to collision with an unstable periodic orbit when a system  
parameter $p$ crosses some critical value $p_c$.  
 
In the dynamics after crisis (for $p \gtrsim p_c$) characteristic  
transients appear; the system spends some time on the former  
(pre-crisis) attractor which is now a chaotic saddle. 
In the case of a {\it boundary crisis} the transient is followed by  
a definitive escape to some other attractor in the phase space 
while after an {\it interior crisis} transients are interrupted by  
(typically short) bursts to an extension of the pre-crisis attractor. 
After {\it attractor merging crisis} we have intermittent jumps  
between symmetric pre-crisis attractors. 
Such a behaviour is called {\it crisis-induced chaos-chaos  
intermittency}. 
For a large class of dynamical systems the time $t$ that the system  
stays on the pre-crisis attractor has an exponential distribution   
$  Pr(t)=\frac{1}{T} \exp(- \frac{t}{T}) $ 
with a mean value $T$ fulfilling a power scaling low  
 
\begin{equation} 
T \sim (p-p_c)^{-\gamma}.    \label{e2} 
\end{equation} 
For two-dimensional maps, the exponent $\gamma>0$ can be expressed  
in terms of the eigenvalues of the  
periodic orbit involved in the crisis \cite{OGY86,b6}.   
The power low (\ref{e2}) and the formulas for the critical exponent 
$\gamma$ \cite{OGY86,b6} have been confirmed in many  
numerical and experimental studies (e.g. \cite{experogy91,rollins84}).
Throughout the recent years many other aspects of crises 
\cite{okryz} as well as new types of crises
\cite{nowekryz} has been investigated.
  
However, it has also been noticed \cite{OGY86,sommerer92}  
that the power low (\ref{e2}) describes only the general tendency of the function 
$T(p-p_c)$, and one can observe some oscillations imposed on it.  
They can result from a ragged (fractal) measure of chaotic attractor 
colliding with its basin of attraction \cite{sommerer91}; in the  
case of homoclinic crisis their period on a log-log plot of  
$T(p-p_c)$ is $|\log|\lambda_2||$, and their amplitude is large for  
small $|\lambda_2|$, where $\lambda_2$ is the contracting eigenvalue  
of the crisis orbit. These oscillations are indicated as a potential  
complication to verifying the scaling low (\ref{e2}) and determining  
the critical exponent $\gamma$. 
 
In this paper we investigate another kind of oscillations that, in  
general, can be more pronounced, and arises from an intertwined  
structure of the basins of attraction \cite{fbbogy85,fbbogy86,greb87}.  
Their particular feature is the existence of sections where the  
average characteristic time $T$ grows up when moving off the crisis  
point, opposite to the general trend (\ref{e2}). Consequently, we call  
them {\it anomalous oscillations}. 
Oscillations of this kind have been observed in a 1D map describing a  
diode resonator \cite{rollins84} and presumably in many other systems  
but, to our knowledge, they were not explicitly pointed out. 
We estimate the period and calculate the amplitude of these  
basin-induced oscillations using a simple model of an attractor  
colliding with its basin boundary.

\section{Crisis in a spin model, anomalous oscillations}
 
First let us briefly introduce a spin model in which the anomalous  
oscillations can be explicitly observed. 
After the papers \cite{b10,kkjh97}, we consider a classical magnetic  
moment (spin) ${\bf S}, |{\bf S}|=S$ in the field of  
uniaxial anisotropy ($z - $ easy/hard axis) with imposed transversal  
magnetic field $\tilde{B}(t)$ along the $x$ axis. The system can be  
described by the Hamiltonian $ H=-A (S_z)^2 - \tilde{B}(t) S_x $,  
where $A$ is the anisotropy constant. Such a model can 
describe a magnetic ion in a paramagnetic material or a single domain 
ferromagnetic sample. 
The motion of the spin is determined by the Landau-Lifschitz equation  
with damping term: 
 
\begin{equation} 
\frac{ d{\bf S} }{dt}= {\bf S} \times {\bf B_{eff} } -  
\frac{\lambda}{S} {\bf S} \times ({\bf S} \times {\bf B_{eff}}),   
\label{e4} 
\end{equation} 
where ${\bf B_{eff}}=-dH/d{\bf S} $ is the effective magnetic field  
and $ \lambda >0  $ is a damping parameter.  
 
Taking the driving field in the form of periodic delta pulses of   
amplitude $B$ and period $\tau$:  
$\tilde{B}(t)= B \sum_{n=1}^{\infty} \delta(t-n\tau) $, 
and using the fact that $|{\bf S}|$ is constant, the equation of  
motion (\ref{e4}) can be transformed into a superposition of two 2D  
maps \cite{b10,kkjh97}: $T_A$ describing the time evolution between  
kicks and $T_B$ describing the effect of the kick. 
The complete dynamics is yielded as a composition of the two maps: 
 
\begin{equation} 
 [S_z^{'}, \varphi^{'}] = T_B[T_A[S_z,\varphi]],    \label{e8} 
\end{equation} 
where $\varphi$ is the angle between the axis $x$ and the projection  
of the spin on the $xy$ plane. 
For different values of the parameters the system exhibits various  
types of dynamics including the periodic and chaotic ones  
\cite{b10,kkjh97} (see also \cite{b12}).  
 
As an example, consider the crisis that occurs at $\tau_c = 2\pi,  
\lambda_c = 0.1054942, A_c=1$ and $B_c=1$ in  
which two symmetric chaotic attractors merge \cite{b10,kkjh97}.  
The attractors  
correspond to two Ising states (spin ``up'' and ``down'') existing in 
the absence of the external field. We take the amplitude 
of the driving field $B$ as an accessible system parameter. For  
$B>B_c$ random jumps between the two, previously separate attractors  
can be observed;  
 
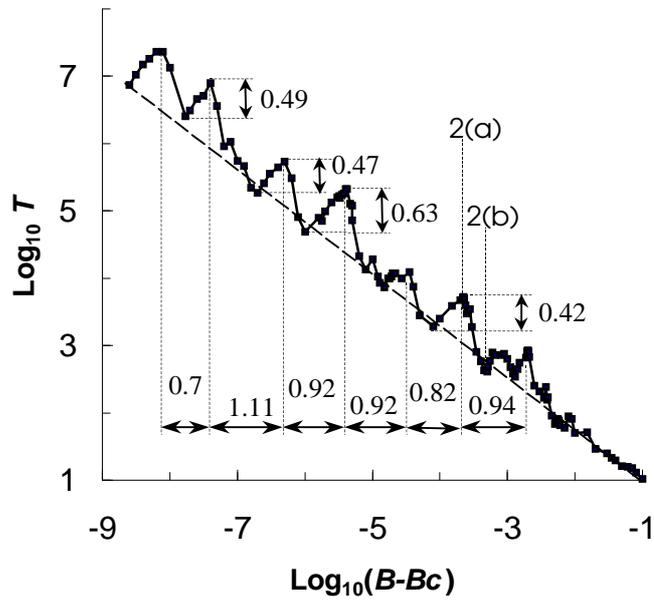
\begin{figure} 
\caption{ The average time of residence on one of the pre-crisis 
attractors versus the distance from the crisis point $A_c=1,  
\tau_c= 2\pi ,\lambda_c=0.1054942..., B_c=1$ for the spin map  
(\ref{e8}).  
Standard deviation is of order or below the size of the 
plotted points. 
Two parameter values which correspond to Fig.~\ref{fig2} are marked. } 
\label{fig1} 
\end{figure}

Fig.~\ref{fig1} shows the dependence of the average time $T$  
between two subsequent switches on the distance $B-B_c$ from the  
crisis point in a log-log scale. We observe a linear trend according 
to (\ref{e2}) with the exponent $\gamma \approx 0.77$ (dashed  
line) \cite{kkjh97}. However, a remarkable, roughly  
periodic oscillations around the trend line can also be seen. These  
oscillations include {\it anomalous} sections where $T$ increases when  
moving 
away from the crisis point. 
 
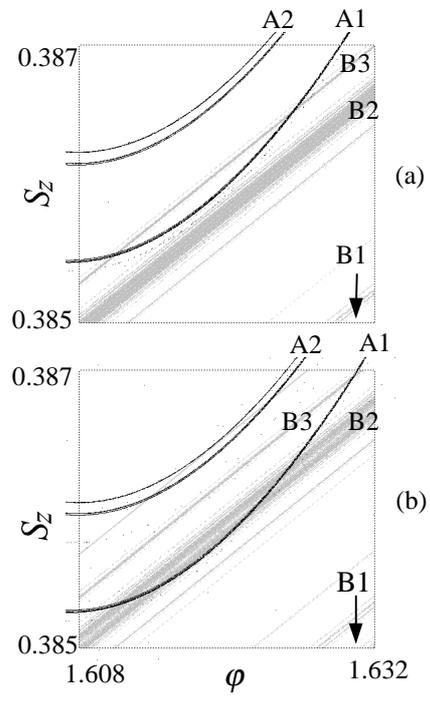
\begin{figure} 
\caption{ Attractor of the spin map (\ref{e8}) creeping into the  
pseudo-basin  
of attraction of the other, symmetric attractor at $A=1, \tau=2\pi, 
\lambda=0,1054942...$ and (a) - $B=1,0000032$, (b) - $B=1,000015$. 
$Am$  ($m=1,2,...$) denotes branches of the attractor, and $Bm$ bands  
of the basin of attraction. } 
\label{fig2} 
\end{figure}

The oscillations are due to an intertwined, fractal structure  
of the basins of attraction of the symmetric pre-crisis attractors. 
To see this, consider the situation depicted in Fig.~\ref{fig2}  
where a part of attractor is plotted together with pseudo-basin  
of attraction (gray spots) of the other, symmetric attractor. 
In fact, after crisis there is only one common attractor and almost 
all points of the phase space form its basin, but for $B \gtrsim B_c$  
we can define {\it pseudo-basins} as sets of initial conditions  
evolving to the ``upper'' ($S_z>0$) or the ``lower'' ($S_z<0$)  
attractor respectively after $M$ iterations, where $M \ll T$;  
in Fig.~\ref{fig2} we took $M=20$. The structure of pseudo-basins 
is much similar to the real basins before crisis. 
In what follows, we shall refer to the post-crisis pseudo-basin 
of attraction of the other attractor as the basin of escape.  
The average transient time $T$ is proportional to the inverse of the 
measure $\mu$ of the part of attractor overlapping with the basin of  
escape \cite{b6}.  
 
In Fig.~\ref{fig2}~(a) the branch $A1$ of the attractor crossed the  
band $B3$ of the basin of escape and is just before the next band  
$B2$. This corresponds to a local maximum in Fig.~\ref{fig1}. 
When the parameter $B$ is increased $A1$ enters the band $B2$,  
so $\mu$ grows and 
$T$ decreases until $A1$ touches the lower edge of $B2$  
(Fig.~\ref{fig2}~(b)). 
Then the total overlap $\mu$ begins to {\it decrease} and we observe an 
anomalous section in Fig.~\ref{fig1}. But simultaneously another branch 
$A2$ enters the basin of escape, so the anomalous section is not as  
pronounced as in the ``clean'' case when $A1$ crosses $B3$. 
 
If the band $B3$ is magnified by a factor of $0.125^{-1}$ a structure  
similar to that on Fig.~\ref{fig2} could be observed; the  
procedure could be repeated further on. This explains the fact that  
the mean period of oscillations in Fig.~\ref{fig1} is approximately  
$|\log_{10}0.125| \approx 0.9$.

\section{Simple model of anomalous oscillations}

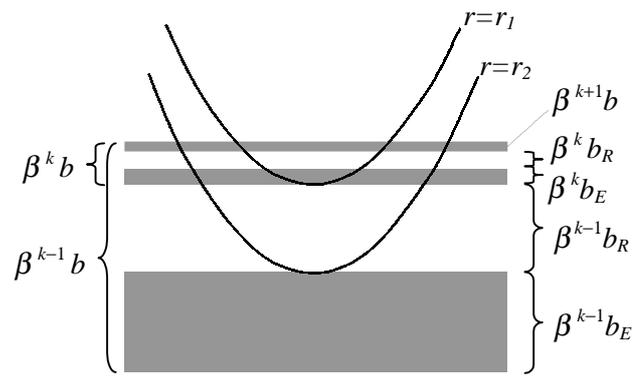
\begin{figure} 
\caption{ Model of an intertwined basin boundary.  } 
\label{fig3} 
\end{figure}

Motivated by the observed self-similarity we introduce  
a simple model of an attractor creeping into an intertwined basin of  
escape in order to assess the amplitude of anomalous oscillations.  
We define the basin in the vicinity of a collision point  
as a self-similar set $\frak{B}$ of stripes of the width $\beta^i b_E$  
accumulating at the line $y=0$ (see Fig.~\ref{fig3}): 
 
\begin{equation} 
{\frak{B}}  =  \bigcup_{i=0}^{\infty} \{ \{x,y\}: y> -\beta^i  b   
\wedge y<-\beta^i  b + \beta^i  b_E \}. 
\label{frba} 
\end{equation} 
$\beta^i b_R$ is the width of the gaps in the basin of escape that 
may be called ``return regions''. The parameters $ b, \beta, b_E $ and  
$b_R$ fulfil the condition $b_E+b_R+\beta b=b$. 
 
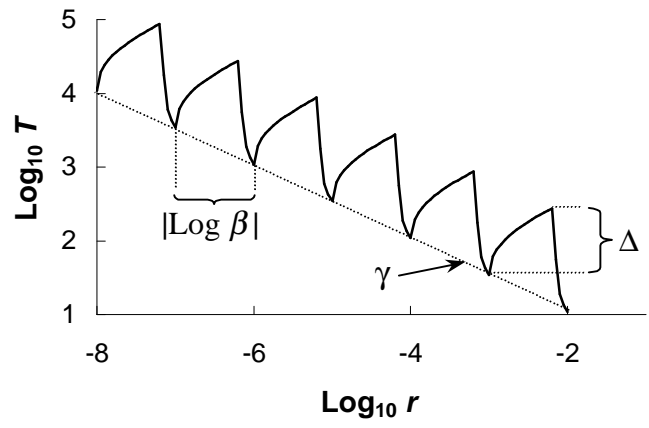
\begin{figure} 
\caption{ Function $T(r)$ obtained when a model one-parabola  
attractor enters the model pseudo-basin of attraction $\frak{B}$ after 
crisis. 
 } 
\label{fig4} 
\end{figure}

As a model attractor we take a single parabola $y=x^2 -r$.  
The parameter $r$ corresponds to the distance $p-p_c$ from the crisis  
point in (\ref{e2}). 
At $r=0$ the attractor collides with the basin $\frak{B}$ (crisis  
occurs). For $r>0$ there is a nonzero overlap of both sets.  
We assume the average transient  
time $T(r)$  proportional to the inverse of the measure $\mu(r)$ of  
the overlapping region (cf. \cite{b6}). Fig.~\ref{fig4} shows the  
function $T(r)$ in log-log scale obtained from the described model.  
Due to self similarity of the model basin $\frak{B}$, we observe  
regular oscillations with the period $\log\beta$ superimposed on the  
linear trend described by (\ref{e2}).  
 
Our aim is  to calculate the amplitude $\Delta$ of the anomalous  
oscillations measured as the difference between a maximum and the 
preceding minimum, see Fig.~\ref{fig4}. 
Consider one $k$-th period of the oscillation. By our definition   
 
\begin{equation}  
\Delta = \log T(r_2) - \log T(r_1) = \log \frac{\mu(r_1)}{\mu(r_2)}, 
 \label{e421}   \end{equation}  
where $r_1 =\beta^k b$, a $ r_2 =\beta^k b +\beta^{k-1} b_R$  
(cf. Fig.~\ref{fig3}).   
 
For small $r$, the length of parabola in the half-plane $y<0$ is  
proportional to $\sqrt{r}$. Using this approximation the measure of  
the overlap $\mu$ at points $r_1$ and $r_2$ can be written as 
 
\begin{equation}     
\mu(r_1) \sim \sqrt{\beta^k b_E},  \, 
\mu(r_2) \sim \sqrt{\beta^{k-1} b_R + \beta^k b_E}  -  
\sqrt{\beta^{k-1} b_R }. 
\label{miodr242}   \end{equation}  
Here we neglected all the strips with $i>k$ in (\ref{frba}). 
From (\ref{e421}) and (\ref{miodr242}) we get  
 
\begin{equation}     
\Delta = (1/\ln q) \sinh^{-1} \sqrt{RE/\beta }, 
\label{delfbb}   \end{equation}  
where $RE = b_R/b_E$ and $q$ is the base of the logarithm in the  
plot $\log T$ vs. $\log r$ (in this Letter we use $q=10$).  
 
One can also calculate the slope of the linear trend,  
getting $\gamma = \frac{1}{2}$ because of a 1D model-attractor. 
Analytical and numerical calculations considering other  
model-attractors show that the fractal 
structure of the basin of escape does not influence the slope  
$\gamma$.

\section{Numerical examples: spin map and H\'enon map}

The formula (\ref{delfbb}) has been verified by comparing the  
calculated $\Delta$ 
with the amplitudes of anomalous oscillations measured on the log-log  
charts $T(p-p_c)$, obtained from computer simulations of the spin 
map (\ref{e8}). 
We investigated the intermittent dynamics near four different crisis  
points  
of the same kind as the one described above, and we made respective 
plots $T(B-B_c)$, similar to Fig.~\ref{fig1}. 
The results are gathered in Table 1. 
The values of $\beta$ and $RE$ used in (\ref{delfbb}) have been  
measured  
employing a series of consecutive magnifications of the basin  
boundary at $B=B_c$ (similar to Fig.~\ref{fig2}).  
Much the same  
values of $\beta$ can be obtained as $10^{-\bar{\theta}}$,  
where $\bar{\theta}$ 
is the average period of anomalous oscillations measured as in  
Fig.~\ref{fig1}. 
 
Note, that the amplitudes of anomalous oscillations calculated from 
(\ref{delfbb}) are close to the maximal amplitudes observed in 
the simulations of the map (\ref{e8}). 
This can be understood noting that the maximal 
amplitude corresponds to a ``clear'' situation  when a branch of  
attractor which may be  considered as one parabola (its fractal fine  
structure may be ignored) leaves a band of the basin while no other  
branch collides with any other basin band. It is just the case  
considered in our model crisis. Conversely, if one branch of attractor  
leaves a basin band and simultaneously another branch enters the same  
or some other band, both effects can cancel and no pronounced peak  
is observed. 
Thus, (\ref{delfbb}) gives the {\it maximal} amplitude of anomalous  
oscillations. 
 
We also studied a homoclinic boundary crisis that occurs in the  
H\'{e}non map, $x_{n+1}=p-x_{n}^{2} - J y_n,  y_{n+1}=x_n$ at  
$J_c=0.3$ and $p_c=2.124672450...$ \cite{OGY86,sommerer91}.  
If we make the $\log_{10}$-scale plot of $T(p-p_c)$ (here $T$ is  
the length of chaotic transient before the escape to infinity,  
averaged over a set of initial conditions) measuring $T$ with  
appropriate accuracy, and marking subsequent points dense enough  
we can see tiny anomalous peaks with average period 0.97 and average 
amplitude (defined as on Fig.~\ref{fig1}) 0.1. 
These oscillations are superimposed on the linear trend (\ref{e2})  
and, here dominating, oscillations due to the ragged measure of the  
chaotic attractor. 
A few consecutive magnifications of the fractal basin boundary gives  
the model parameter values $RE \approx 0.0167 $ and  
$\beta \approx 0.107$. 
The formula (\ref{delfbb}) then gives $\Delta = 0.168$. This value is  
again close to the maximal amplitude observed ($\approx 0.16$). 
Note, that the average period of anomalous oscillations is  
approximately $|\log_{10} \beta|$.

\section{Discussion and summary}
 
The presented model used to obtain the maximal amplitude of
anomalous oscillations is a simple one. In real basins of attraction
every strip of ${\frak B}$ may have its own fine structure and chaotic 
attractor is locally an infinite fractal set of parabolas. Nevertheless,
the model comprises the main properties of both sets responsible
for the anomalous oscillations and gives a proper estimate of
their maximal amplitude.
 
Anomalous oscillations seem to be quite common in nonlinear  
dynamical systems, as do the fractal, intertwined basin boundaries that 
give rise to them. They are, however, easy to confuse with statistical  
dispersion of the mean times $T$, especially when one has a small  
number of points on $T(p-p_c)$ plots.  
The presence of anomalous oscillations can be a complication to verify  
the scaling low (\ref{e2}), but if one is able to detect and measure  
them, they can give additional information on the structure of  
basins of attraction near crises. Note, that both $\Delta$ and $\beta$  
can be, in principle, assessed from the plot (like in Fig.~\ref{fig1})  
and Eqn. (\ref{delfbb}) can be used to derive $RE$. 
 
A model of chaotic attractor similar to that of the basin of attraction 
lets us calculate the  
amplitude of ``normal'' oscillations caused by the fractal structure 
of the attractor \cite{kkjhtbp}.  
 
Concluding, we have observed anomalous oscillations of average  
characteristic crisis-induced transient times, and explained them  
in terms of a simple, geometric 
model, which gives the period of the oscillations and enables to  
calculate with a fair accuracy their maximal amplitude.  
The model is universal, and the results are applicable 
to a large class of 2D systems undergoing different types of crises.  
It could be also customized for some special non generic cases.

\section{Acknowledgements}  
 
This work has been partly supported by the Polish Committee for  
Scientific Research (KBN), Grant No. 2P03B 031 14. A part of the  
calculations has been performed on CRAY 6400 at Warsaw University  
of Technology.

 
 
\begin{table}[h] 
 
\begin{tabular}{ddd|dd|ddd} 
 \multicolumn{3}{c|}{Crisis parameters} & 
  \multicolumn{2}{c|}{Model parameters} &  
\multicolumn{3}{c}{Amplitude of oscillations}   \\   
\multicolumn{1}{c}{$B$} & \multicolumn{1}{c}{$\tau$} &  
\multicolumn{1}{c|}{$\lambda$} &  
 \multicolumn{1}{c}{$\beta$} & \multicolumn{1}{c|}{$RE$} & 
\multicolumn{1}{c}{$\Delta_{av}$} &  
\multicolumn{1}{c}{$\Delta_{max}$} &  
\multicolumn{1}{c}{$\Delta_{th}$}  \\  \hline 
 1    & $2\pi$ & 0.10549 & 0.125 & 0.5  & 0.502 & 0.63 & 0.627 \\ 
 0.92 & $2\pi$ & 0.08964 & 0.08  & 0.95   & 0.7   & 0.8  & 0.847 \\ 
 0.92 & 5.8866 & 0.08    & 0.16  & 0.15  & 0.325 & 0.37 & 0.373 \\ 
 1.2  & $2\pi$ & 0.1437  & 0.124 & 0.077 & 0.26  & 0.31 & 0.314 \\ 
\end{tabular} 
\caption{Amplitudes of anomalous oscillations of average characteristic  
times $T$ in the vicinity of four different crisis points in 
the spin map (\ref{e8}) at $A=1$. The average ($\Delta_{av}$) and  
maximal ($\Delta_{max}$) amplitudes are compared to the values  
$\Delta_{th}$ 
obtained from (\ref{delfbb}) with respective $\beta$ and $RE$.} 
 
\end{table}

\end{document}